\documentclass{aa}
\usepackage{txfonts}
\usepackage{latexsym}
\usepackage{pictex}
\usepackage{amssymb}
\usepackage{graphics}
\usepackage{afterpage}
\begin{document}

   \title{{\bf VPMS\,J1342+2840} -- an unusual quasar from the variability and proper 
   motion survey
}

   \author{H. Meusinger\inst{1}
           \thanks{
Visiting Astronomer, German-Spanish Astronomical Centre, Calar Alto,
operated by the Max-Planck-Institute for Astronomy, Heidelberg, jointly 
with the Spanish National Commission for Astronomy
                  }\fnmsep 
           \thanks{Visiting astronomer, Italian Telescopio 
Nazionale Galileo (TNG) operated at the Spanish Observatorio 
del Roque de los Muchachos of the Instituto de 
Astrofisica de Canarias
                  }
          \and D. Froebrich\inst{2}$^{\, \star}$
	  \and M. Haas\inst{3}
	  \and M. Irwin\inst{4}
	  \and M. Laget\inst{5}
	  \and R.-D. Scholz\inst{6}
          }

   \institute{
Th\"uringer Landessternwarte Tautenburg, 07778 Tautenburg, Germany,
e-mail: meus@tls-tautenburg.de
\and
Dublin Institute for Advanced Studies, 5 Merrion Square, Dublin 2, Ireland
\and
Astronomisches Institut, Ruhr-Universit\"at Bochum, 
Universit\"atsstr.\,150/NA7, 44780 Bochum, Germany               
\and
Institute of Astronomy, Madingley Road, Cambridge CB3 1HA, UK
\and
Laboratoire d'Astrophysique de Marseille, B.P. 8, F-13376 Marseille cedex
12, France
\and
Astrophysikalisches Institut Potsdam, An der Sternwarte 16, 14482 Potsdam, 
Germany
             }

%   \offprints{H. Meusinger, \email{meus@tls-tautenburg.de}}

   \date{Received ; accepted }

   \abstract{
   We report the discovery of the highly peculiar, radio-loud 
   quasar  VPMS\,J1342+2840 ($z$$\approx$1.3) from the variability and 
   proper motion survey.
   We present spectroscopic, imaging and photometric observations. The 
   unusual spectrum shows a strong depression of the continuum over a wide 
   wavelength range in the blue part without the typical structures of 
   broad absorption line (BAL) troughs. The image of the quasar is unresolved 
   and there is no evidence for a foreground object on the line of sight.
   The broad-band spectral energy distribution is not consistent with obvious
   dust reddening with the standard SMC extinction curve.
   The downturn of the continuum flux of VPMS\,J1342+2840 at short wavelengths 
   can be caused by dust reddening only if the reddening curve is steeper 
   then the SMC curve in the ultraviolet and is very flat at longer wavelengths. 
   Alternatively, the dominant spectral features can be explained by 
   low-ionization BALs forming unusually wide, overlapping absorption troughs.
}

\authorrunning{Meusinger et al.}
\titlerunning{VPMS unusual BAL quasar}

   \maketitle

%%%%%%%%%%%%%%%%%%%%%%%%%%%%%%%%%%%%%%%%%%%%%%%%%%%%%%%%%%%%%%%%%%%%%%%%%%%%%
%
%
%
\section{Introduction}
%
%
%%%%%%%%%%%%%%%%%%%%%%%%%%%%%%%%%%%%%%%%%%%%%%%%%%%%%%%%%%%%%%%%%%%%%%%%%%%%%

An important class of active galactic nuclei are quasars with broad absorption 
lines (BAL), usually subdivided into low- and 
high-ionization (LoBAL, HiBAL) quasars. 
BAL quasars constitute about 
15\% of the quasar population (Reichard et al. \cite{Rei03})
and only about 10\% of the BAL quasars are LoBALs. 
A rare subclass of the LoBALs are the iron LoBALs (FeLoBALs) with 
absorption from metastable excited states of \ion{Fe}{II}\, and \ion{Fe}{III}\, 
(Hazard et al. \cite{Haz87}). 
BALs are naturally explained by powerful,  non-isotropic, subrelativistic 
outflows (Weymann et al. \cite{Wey91}), possibly 
related to disk winds (Murray \& Chiang \cite{Mur98}). 
LoBALs probably have higher gas column densities
and at least some LoBAL quasars 
are thus expected to represent either young systems expelling a thick shroud 
of gas and dust (Voit et al. \cite{Voi93}; Canalizo \& Stockton \cite{Can01}) 
or a  transition stage between radio-loud and radio-quiet BAL quasars 
(Becker et al. \cite{Bec97}). 

Quasars with unconventional spectra are obviously 
overlooked in quasar surveys based on optical/UV colour or objective prism
selection. They are also under-represented in radio surveys with high flux 
density limits because only a small percentage seems to be radio-loud.  
Unusual BAL quasars from the 
Sloan Digital Sky Survey (SDSS) First Data Release were analyzed in detail 
by Hall  et al. (\cite{Hal02}; hereafter H02) and the most unconventional quasars from the 
SDSS Second Data Release  were presented by Hall et al. (\cite{Hal04a}a).
Menou et al. (\cite{Men01}) discuss SDSS BAL quasars with radio detection 
in the VLA FIRST survey. Other quasars with unconventional spectra, 
e.g. unusual reddening in the UV (Hines \cite{Hin01}) or without 
obvious metal-line emission (Hall et al. \cite{Hal04b}b), 
have been the subject of a variety of studies. The data base is clearly 
growing, nevertheless every single odd quasar is an interesting object 
that may hold clues to various open questions of 
the structure and the evolution of quasars.
Here, we present one of the 
most unusual quasars from the variability and proper motion survey (VPMS). 
The VPMS is an optical quasar 
search project based upon long-term variability and zero proper motion 
(Scholz et al. \cite{Sch97}; Meusinger et al. \cite{Meu02}, \cite{Meu03}). 
The quasar selection does not explicitly make use of assumptions on the 
spectral energy distribution (SED) of quasars and thus the VPMS is expected 
to be less biased against quasars with unusual SEDs than more conventional 
optical quasar surveys.

%%%%%%%%%%%%%%%%%%%%%%%%%%%%%%%%%%%%%%%%%%%%%%%%%%%%%%%%%%%%%%%%%%%%%%%%%%%%%
% 
%
%
\section{Spectroscopy and imaging}
%
%
%%%%%%%%%%%%%%%%%%%%%%%%%%%%%%%%%%%%%%%%%%%%%%%%%%%%%%%%%%%%%%%%%%%%%%%%%%%%%

\object{VPMS\,J1342+2840} 
($13^{\rm h}42^{\rm m}46\fs25$\,+28$\degr 40\arcmin 27\farcs5$,\, J2000.0)
is a faint ($\bar{B}=20.5$) VPMS quasar, discovered during 
regular spectroscopic follow-up observations of high-priority  
quasar candidates in spring 2003. Subsequently, 
several spectra of moderately higher resolution 
 were taken with CAFOS at the 2.2\,m telescope on 
Calar Alto, Spain, and with DOLORES at the 3.6\,m Telescopio Nazionale Galileo 
(TNG) on the island of La Palma, Spain, respectively (Table\,1).  
All observations were done 
in good atmospheric conditions.
The spectra were reduced using the optimal extraction 
algorithm of Horne (\cite{Hor86}).

\begin{table}[htbp]
\caption{Summary of spectroscopic observations. Column 4 gives the resolution
expressed by FWHM.}
\begin{tabular}{cccccc}
\hline\hline
instrument  & grism & $\lambda$ range &  FWHM  & $t_{\rm exp}$     & epoch\\
            &       & (\AA)           & (\AA)  & (ks)              & (year)\\
\hline
  CAFOS   & B400 & 3\,600-8\,000   &  34	 & 3.9  	   & 2003.3 \\
  CAFOS   & R200 & 6\,300-11\,000  &  19	 & 1.2  	   & 2003.6 \\
  CAFOS   & G200 & 5\,200-9\,000   &  15	 & 5.4  	   & 2004.6 \\
DOLORES   & LR-B & 3\,600-8\,000   &  22	 & 1.8  	   & 2004.7 \\
DOLORES   & LR-R & 5\,200-9\,000   &  15	 & 1.2  	   & 2004.7 \\
\hline
\end{tabular}
\label{mags}
\end{table}

\begin{figure}[thtp]   %Fig.1
\vspace{6.5cm}
\includegraphics{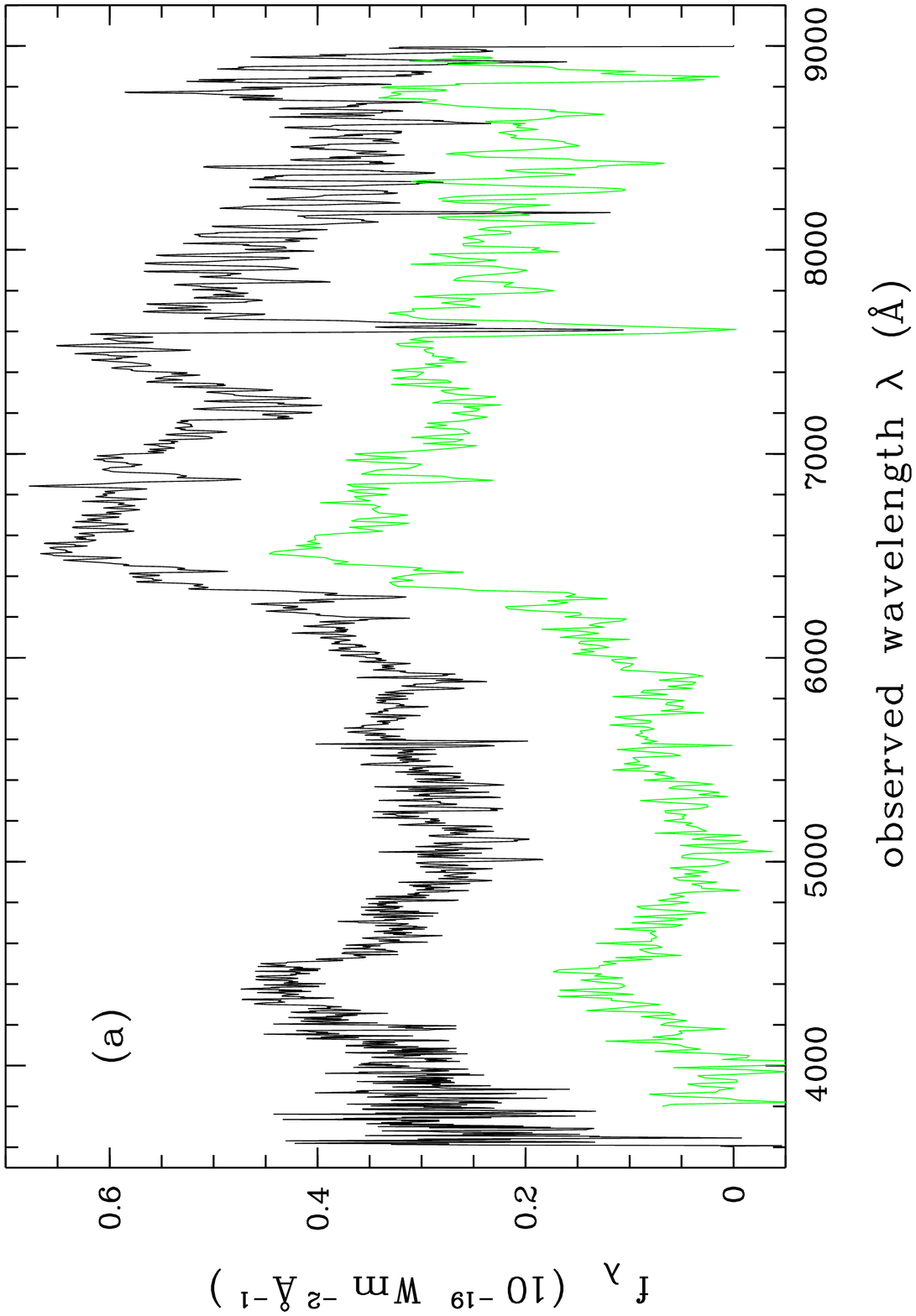}
\vspace{6.2cm}		
\includegraphics{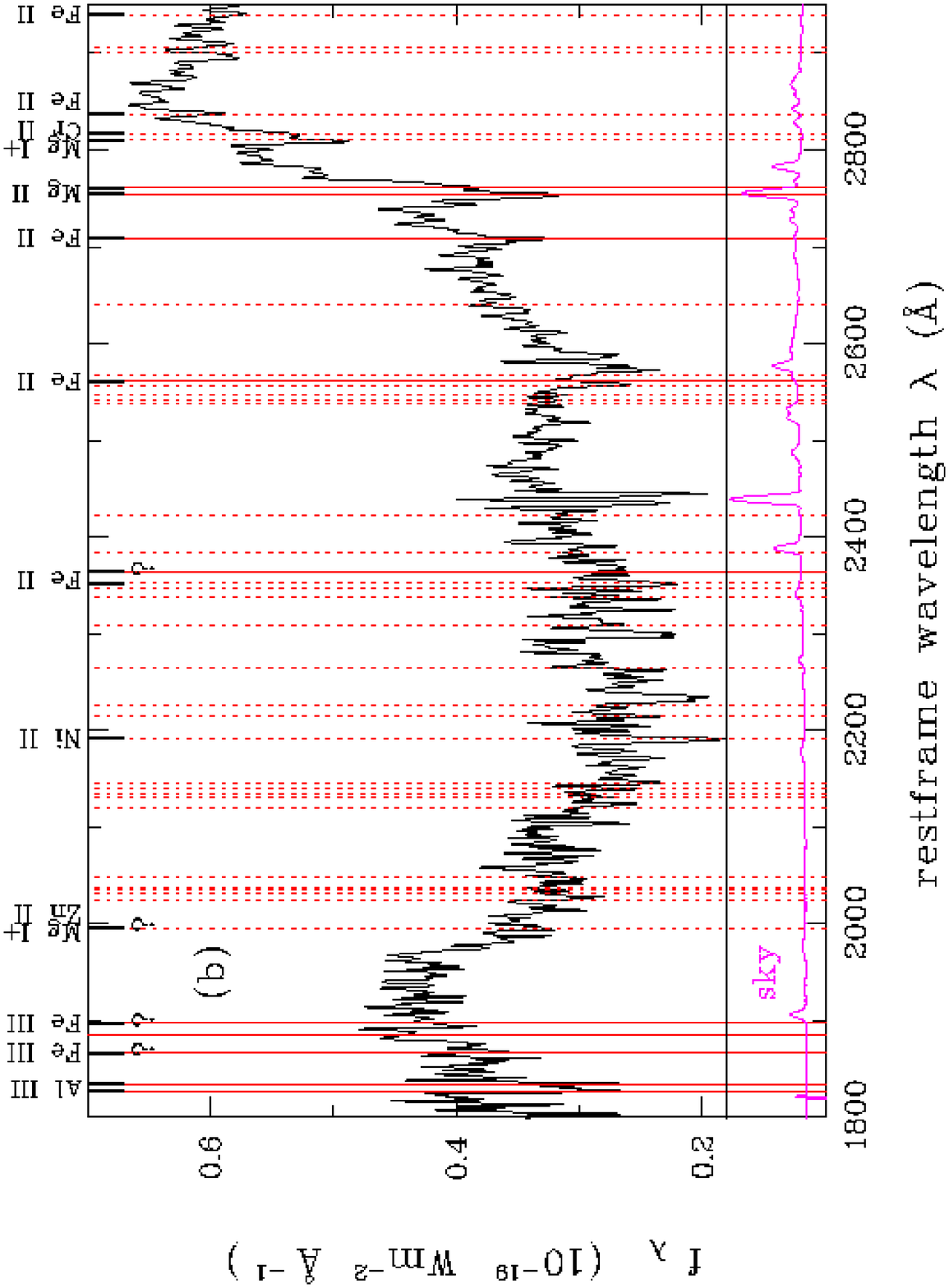}
 \caption{{\bf a)} 
 TNG spectrum in the observer frame (dark curve). For 
 comparison, the lower-resolution CAFOS (B400+R200) spectrum is also shown, 
 shifted arbitrarily downward for lucidity. The spectra are not corrected for 
 telluric absorption lines.
 {\bf b)} The region between 1800 and 3000\AA\, of the TNG spectrum 
 in the quasar emission frame. The CAFOS G200 spectrum is similar
 but has a lower signal-to-noise and is not shown here.
 Vertical lines: firmly identified 
 transitions in BAL quasars from H02 (their table\,1); solid lines
 for transitions that are common in LoBAL and FeLoBAL quasars, 
 dashed lines for rare or unusual transitions. Identified lines  
 are marked at the top (``?'' for uncertain identification). 
 Bottom: night sky spectrum (not to scale).
    }
  \label{spec}
\end{figure}

With its strong depression of the continuum over the whole wavelength
interval shortward of 6500\AA\, (Fig.\,\ref{spec}a), the spectrum of 
VPMS\,J1342+2840 is remarkably different from the typical quasar 
spectrum (see Fig.\,\ref{sed}a). 
It is tempting to interpret this depression as due to strong absorption 
from a blend of iron lines in the spectral region 
$\lambda \approx$2300-3050\AA\, and 
to identify the bump at $\lambda$\,4400\,\AA\, with emission from 
\ion{C}{III}]\,$\lambda1909$ 
corresponding to an (uncertain) systemic redshift of 
$z_{\rm em}$$\approx$1.288 (i.e., $M_{\rm B}$$\approx$$-27$, after correction 
for absorption).  This interpretation is supported by the 
identification of several narrow absorption lines (Fig.\,\ref{spec}b) 
at $z_{\rm abs}$=1.2535, most notably
\ion{Al}{III}\,$\lambda$\,1860,
\ion{Ni}{II}(UV12,UV13)\,$\lambda$\,2225,
\ion{Fe}{II}(UV1)\,$\lambda$\,2600,
\ion{Fe}{II}(UV62,UV63)\,$\lambda$\,2750,
\ion{Mg}{II}\,$\lambda$\,2800,
\ion{Mg}{I}\,$\lambda$\,2853,
\ion{Fe}{II}(UV61)\,$\lambda$\,2880.
The \ion{Fe}{II}\,$\lambda$\,2750 line seems to be unphysically strong 
compared to the iron lines at 2400 and 2600\,\AA. The comparison 
with the CAFOS G200 spectrum shows that the line is possibly slightly enhanced 
by noise in the TNG spectrum. Nevertheless, the detection of \ion{Fe}{II} lines 
is beyond doubt. There might be also other absorption line systems at 
slightly different redshifts, yet evidence for them is not 
as clear as for the $z_{\rm abs}$=1.2535 system.
The abrupt drop-off in flux near \ion{Mg}{ii}\,$\lambda$\,2800 
is of course not explained by one narrow-line system.
Such a drop-off is known from several
FeLoBAL quasars. However, there are no obvious BAL troughs in 
VPMS\,J1342+2840. In this respect, 
VPMS\,J1342+2840 resembles SDSS\,J0105-0033 and SDSS\,J2204+0031, 
which were classified by H02 as  ``mysterious objects''.

%\vspace{-0.5cm}
%----------------------------------------------------------------
%----------------------------------------------------------------

%%%%%%%%%%%%%%%%%%%%%%%%%%%%%%%%%%%%%%%%%%%%%%%%%%%%%%%%%%%%%%%%%%%%%%%%%%%%%
%
%
%
%\section{Imaging}
%
%
%%%%%%%%%%%%%%%%%%%%%%%%%%%%%%%%%%%%%%%%%%%%%%%%%%%%%%%%%%%%%%%%%%%%%%%%%%%%%

\begin{figure}[htbp]  % Fig.2
%\fbox{\resizebox{8.6cm}{8.6cm}{\includegraphics{Gl232_f2_neu.eps}}}
\vspace{8.8cm}
\includegraphics{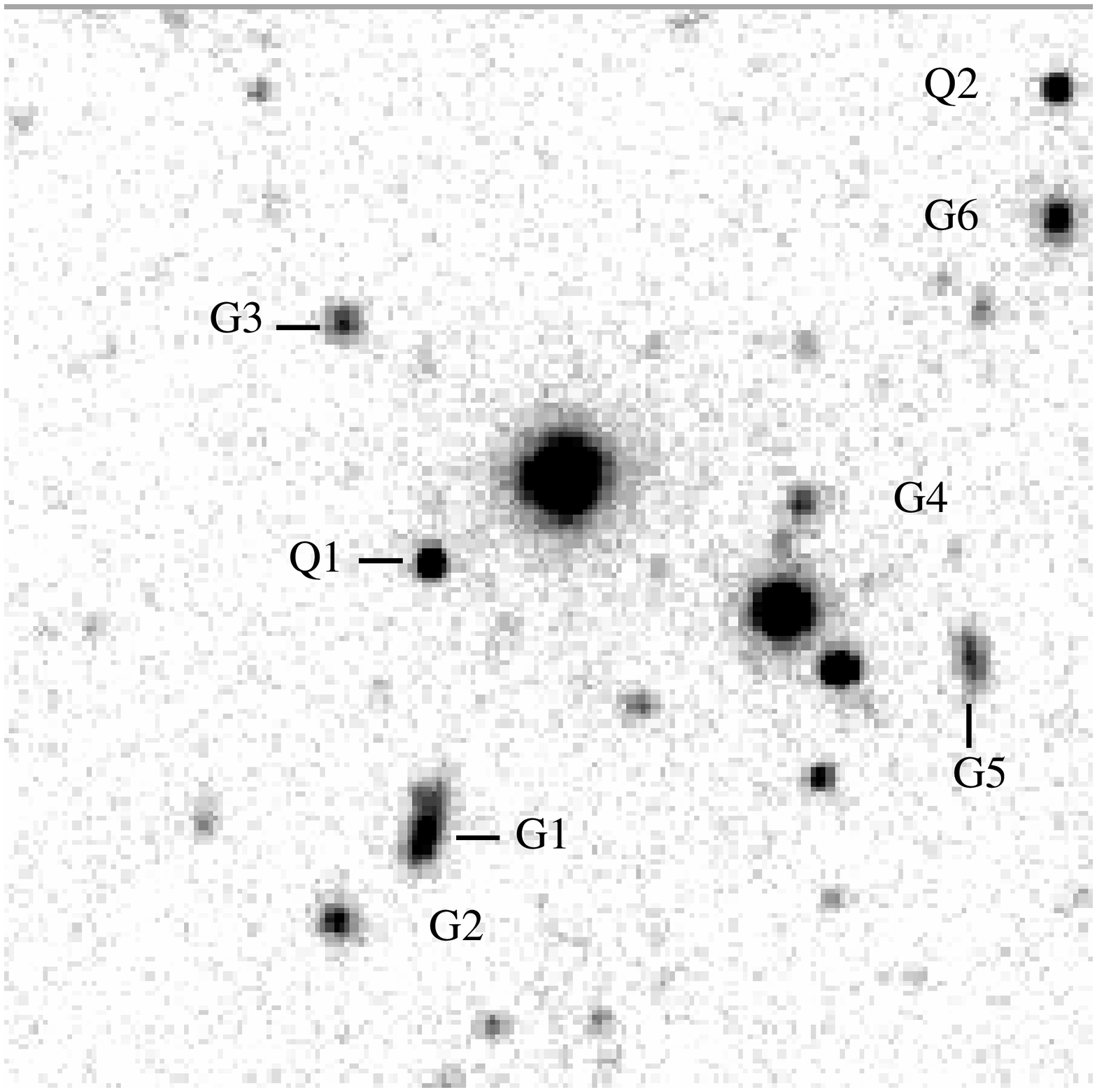}
\includegraphics{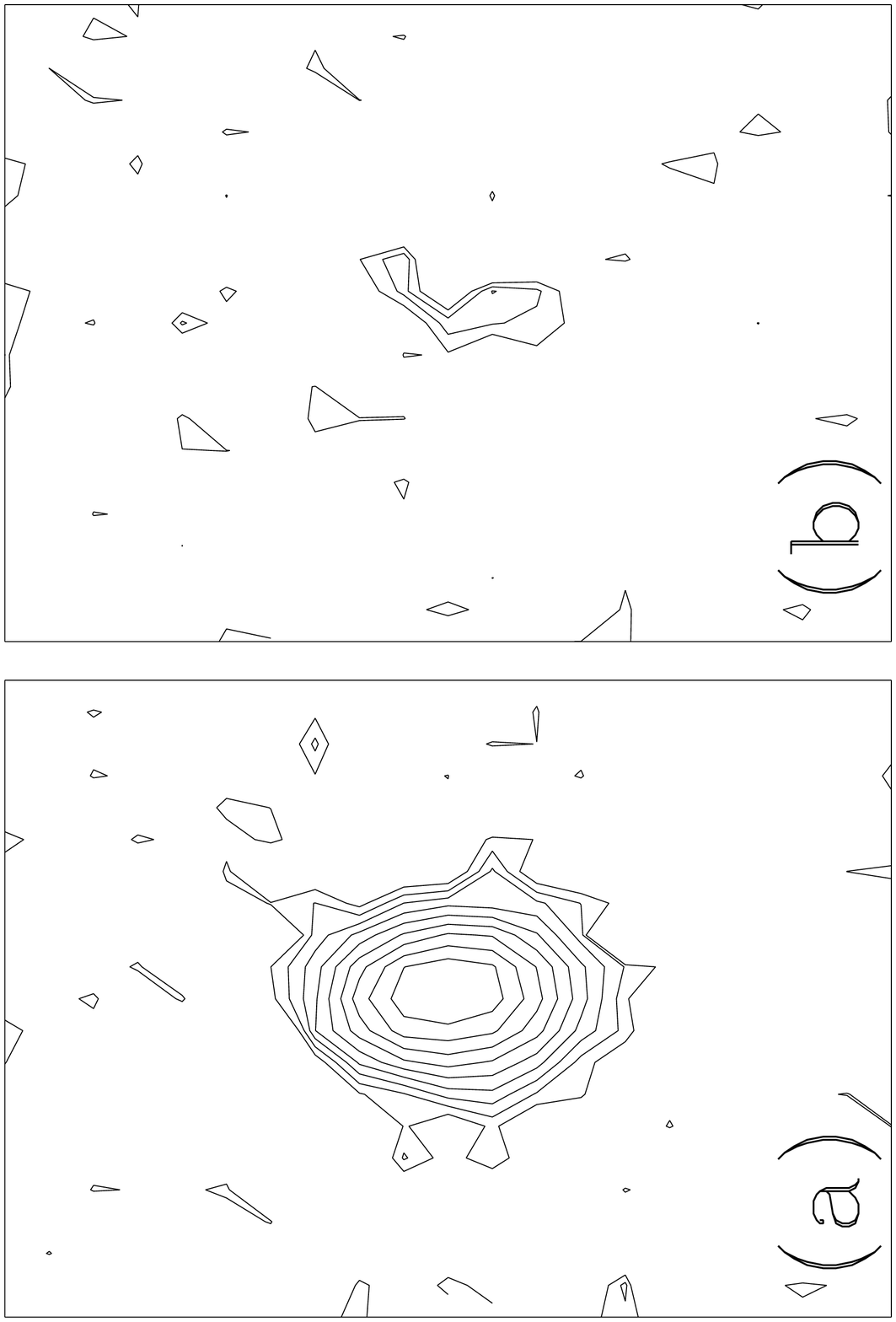}
\caption{$90\arcsec \times 90\arcsec$ field around VPMS\,J1342+2840 (Q1)
 in the R band (N is up, E is left).  
 The other quasar (Q2) in the image is VPMS\,J1342+2841 ($z=0.873$).
 Larger extended objects are labeled with ``G''. The bright star in the center
 is GSC2.2\,N1301033185 ($R$=14.9). 
 {\it Inset:} contour plots 
 (equally spaced logarithmic intervals) of the 10\arcsec$\times10$\arcsec\, 
 area containing {\bf a)} the original image of  VPMS\,J1342+2840 and 
 {\bf b)} the residuals  after subtraction of the PSF. 
}
\label{optical}
\end{figure}

Unusual spectral properties can be the result of a positional coincidence
with a foreground object  in combination with gravitational 
lensing (Irwin et al. \cite{Irw98}; Chartas \cite{Cha00}).
A deep ($R_{\rm lim}$$\approx$23) R band image taken with CAFOS
(Fig.\,\ref{optical}) reveals several faint galaxies within $\sim 1\arcmin$ 
and faint, fuzzy structures within $\sim 10\arcsec$. 
Granted that the faintest 
structures are associated with the quasar environment, their absolute 
magnitudes would be $M_{\rm R}$$\approx$$-24.5$, 
i.e. $\sim$2\ldots3.5\,mag brighter than the Schechter magnitude
$M_{\rm R}^{\ast}$ 
(e.g., Gaidos \cite{Gai97}; Chen et al. \cite{Che03}; 
Christlein \& Zabludoff \cite{Chr03}; 
$H_{\rm 0}$=65\,km\,s$^{-1}$\,Mpc$^{-1},\, q_{\rm 0}$=0.5,\,$\Lambda$=0).
Such bright galaxies are extremely rare and thus we conclude that the 
``fuzz'' is more likely to represent foreground galaxies.
For the galaxy G1 we measured a spectroscopic redshift of $z$=0.142.
However, there is no indication for foreground absorption at this 
redshift in the spectrum of VPMS\,J1342+2840. Furthermore,
the image of VPMS\,J1342+2840 is unresolved; there is no indication for a  superposition 
with another object (see inset of Fig.\ref{optical}). 
% from the analysis of the quasar image. The inset of Fig.\ref{optical} shows the 
% contour plot of VPMS\,J1342+2840 {\bf a)} before and {\bf b)} after 
% subtraction of the PSF. 
The faint residuals in the PSF-subtracted image 
are comparable with the residuals typically found for the PSF-stars. 
Finally, the photometric variability of VPMS\,J1342+2840 (see below) is
common in quasars but is inexplicable for a lensed background galaxy.

%----------------------------------------------------------------
%----------------------------------------------------------------

%%%%%%%%%%%%%%%%%%%%%%%%%%%%%%%%%%%%%%%%%%%%%%%%%%%%%%%%%%%%%%%%%%%%%%%%%%%%%
%
%
%
\section{Broad-band spectral energy distribution}
%
%
%%%%%%%%%%%%%%%%%%%%%%%%%%%%%%%%%%%%%%%%%%%%%%%%%%%%%%%%%%%%%%%%%%%%%%%%%%%%%

To study the SED over a broader spectral range we consider photometric data 
from several bands: photometry in the JHK$'$ bands was performed with MAGIC at 
the 2.2\,m telescope on Calar Alto.  BVR magnitudes are from
images taken with CAFOS at the same telescope. The U magnitude
comes from the VPMS measurements on Schmidt plates taken later than 1990
(see below).  
The field has been observed with the balloon-borne ultraviolet 
imaging telescope FOCA (Laget et al. \cite{Lag92}) at $\lambda$\,2000\,\AA;
the flux limit at half completeness is 0.026\,mJy. As expected,
there is no significant detection of VPMS\,J1342+2840 
since the Lyman limit comes close to the FOCA band for 
$z \approx 1.3$. Finally, 
an I magnitude, without error bar, is available 
from the USNO-B1.0 catalogue (Monet et al. \cite{Mon03}) measured on a 
POSS\,II {\sc iv}-N plate taken in 1997. 

\begin{figure}[bhtp]   %Fig.3
\vspace{6.2cm}
\includegraphics{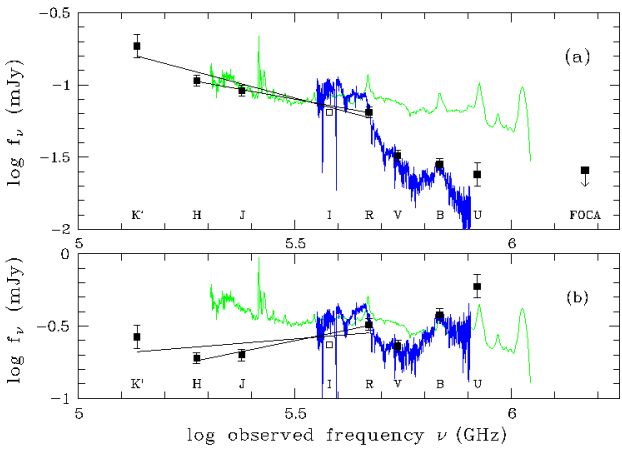}
 \caption{{\bf a)} 
 The observed SED from photometric  data in various bands (symbols)
 with linear regression lines (see text). Superimposed are the 
 TNG spectrum from Fig.\,\ref{spec} (dark curve) and the mean VPMS 
 quasar spectrum (thin curve). The fluxes were corrected for 
 galactic foreground absorption using the extinction maps of Schlegel et al. 
 (\cite{Sch98}). 
 {\bf b)} The same after dust reddening correction for the SMC extinction 
 curve from Pei (\cite{Pei92}) with $E(B-V)$=0.3. 
     }
  \label{sed}
\end{figure}

The resulting broad-band SED is shown in Fig.\,\ref{sed}a.
The uncertainty due to variability is expected to be small.
However, we caution that changes in flux cannot be ruled out.
The B band lightcurve from the VPMS data shows a 
dimming by $\sim$0.8\,mag between $\sim$1970 and $\sim$1990  but 
only small fluctuations between $\sim$1990 and 2003. 
A similar behaviour is indicated by the UVR data. (The dimming is stronger  
at shorter wavelengths, as is typical for quasars, cf. 
Tr\'evese et al. \cite{Tre01}).
Fig.\,\ref{sed} reflects the depression of the continuum 
seen in Fig.\ref{spec}\,a. We restrict the power-law approximation 
$f_{\nu}\propto\nu^{\,\alpha}$ to $\lambda$$>$$6500$\,\AA\, (observed) 
and find $\alpha$=$-0.80$ from the linear regression of the K$'$HJR 
magnitudes. If we exclude the K$'$ data point 
(it is not clear whether the turnup at 2.2\,$\mu$m is real), 
the fit is considerably improved and the slope $\alpha$=$-0.54$ 
comes close to the slope of the composite spectrum from all VPMS quasars.

We found a positionally coincident ($\sim$\,1\arcsec) 
radio source on the image cutout from the NRAO Very Large Array FIRST survey 
(Becker et al. \cite{Bec95}) with a  peak flux density of 0.7\,mJy and an 
integrated flux density of 2.3\,mJy at 1.4\,GHz. 
The source is not listed in the FIRST catalogues because of the hard detection
limit of 1\,mJy, which applies to the peak flux density rather than to the
integrated flux density.  
In the NRAO 1.4\,GHz VLA Sky Survey (NVSS; Condon et al. \cite{Con98}),
a source with a peak flux density of 1.6\,mJy is seen at a distance of 
$\sim$\,5\arcsec. 
According to NED\footnote{The NASA/IPAC Extragalactic Database is operated by
the Jet Propulsion Laboratory, California Institute of Technology, under 
contract to NASA.}, 
VPMS\,J1342+2840 is not detected at other radio bands; thus the radio spectral
index is not available. We use the K-corrected 
ratio $R^{\ast}$ of the 5\,GHz radio flux density to 
the 2500\,\AA\, optical flux density in the quasar rest frame as a measure 
of radio-loudness. Using the standard definition with 
$\alpha_{\rm radio}$=$-0.5$ and $\alpha_{\rm opt}$=$-1$  
(Stocke et al. \cite{Sto92}), 
we find $\log\,R^{\ast}$=$1.7$. If we consider that 
{\it (a)} the observed flux at 2500\,\AA\ (rest frame) is reduced and  
{\it (b)}  $\alpha_{\rm opt}$=$-0.54$ or $-0.80$ (Fig.\,\ref{sed}a), 
we get $\log\,R^{\ast}$=$1.52$ or 1.56, respectively, 
for the FIRST radio flux.
With $\alpha_{\rm radio}$=$-0.3$ (White et al. \cite{Whi00}),
the corresponding values  are $\log\,R^{\ast}$=$1.55$ and 1.60.
Adopting a divide between radio-quiet and radio-loud quasars 
at $\log\,R^{\ast}$$\approx$1 (e.g., Stocke et al. \cite{Sto92}),
VPMS\,J1342+2840 has to be classified as radio-loud.

%----------------------------------------------------------------
%----------------------------------------------------------------

%%%%%%%%%%%%%%%%%%%%%%%%%%%%%%%%%%%%%%%%%%%%%%%%%%%%%%%%%%%%%%%%%%%%%%%%%%%%%
%
%
%
\section{Discussion}
%
%
%%%%%%%%%%%%%%%%%%%%%%%%%%%%%%%%%%%%%%%%%%%%%%%%%%%%%%%%%%%%%%%%%%%%%%%%%%%%%

Is the peculiar spectrum of VPMS\,J1342+2840 due to strong dust
extinction?  The reddening associated with mildly obscured quasars has 
typically been found to be consistent with that of the
Small Magellanic Cloud (SMC; Hopkins et al. \cite{Hop04}).
On the other hand, Gaskell et al (\cite{Gas04}) argue that most 
quasars are affected by extinction with a reddening curve which 
is distinct from the SMC-curve. 
To de-redden the spectrum of VPMS\,J1342+2840 we adopt three different
reddening curves: 
{\it (a)} SMC-like, 
{\it (b)} relation from Gaskell et al.,  
{\it (c)} Galactic curve.
The colour excess $E(B-V)$ is used as a free parameter to 
fit the blue end of the corrected TNG spectrum to the composite 
spectrum from all VPMS quasars which was normalized before to 
the red end of the corrected TNG spectrum.
For SMC-like reddening, the de-reddened blue ($\lambda$$\la$2000\,\AA) 
end of the corrected continuum matches the mean quasar spectrum
for $E(B-V)\approx$0.3 (Fig.\,\ref{sed}b). However, there is a strong 
discrepancy in the intermediate 
part of the spectrum (2000 to 2800\,\AA\, rest frame). 
Moreover, the de-reddened broad-band SED has a very untypical shape 
and does not fit the mean quasar spectrum over a broader spectral range.  
The fit is even worse for Galactic extinction with its 
$\lambda$\,2175\AA\, bump and it is much worse for the curve from Gaskell et al. 
which is too flat at short wavelengths.  
Given that the slope of the 
red and near-IR part of the uncorrected broad-band SED is well matched 
by the usual power-law, we conclude that dust can be a 
major reason for the peculiar spectrum of VPMS\,J1342+2840
only for an unusual extinction curve which must be very flat at
$\lambda$$\ga$2900\AA\, and steeply rising at shorter wavelengths.
Stronger than SMC-like reddening in the UV has been suggested
by H02. Hines et al. (\cite{Hin01}) argue that the downturn at 
$\lambda$$\la$2500\AA\, in the spectra of two IRAS 
quasars might be produced by dusty scattering.

Another likely explanation for the depression at $\lambda \la 3000$\AA\,
(rest frame) are wide, overlapping BAL troughs.
Such an interpretation requires absorption
that increases in strength when BALs overlap. This problem has been 
discussed at length by H02 in the context of the 
two ``mysterious objects'' 
SDSS\,J0105-0033 and SDSS\,J2204+0031. 
These authors argue that 
barely plausible fits can be achieved 
with wide troughs detached by $\sim$12\,000\,km\,s$^{-1}$ 
if there is partial covering of different regions of the 
continuum source as a function of velocity. 
Partial covering seems not implausible for VPMS\,J1342+2840: it is known that absorption in BAL 
outflows is saturated, even though the troughs rarely reach zero flux. 
This nonblack saturation can be due to scattered light that bypasses 
the absorbing region or due to partial covering of the continuum source 
(Arav et al. \cite{Ara99}). Because of the strong variability 
on timescales of a few years it seems unlikely that the observed blue
continuum of VPMS\,J1342+2840 is dominated by scattered light. 
Following H02, the only object known to definitely exhibit spatially 
distinct velocity-dependent partial covering is FBQS\,1408+3054. 
If our interpretation is true,  VPMS\,J1342+2840 would be another 
representative of these extremely rare objects. Moreover,
it is tempting to speculate that VPMS\,J1342+2840 is a
``missing link''  (perhaps transitional) between FBQS\,1408+3054 
and the ``mysterious'' SDSS quasars.

%%%%%%%%%%%%%%%%%%%%%%%%%%%%%%%%%%%%%%%%%%%%%%%%%%%%%%%%%%%%%%%%%%%%%%%%%%%%%
%
%
%
\section{Conclusion}
%
%
%%%%%%%%%%%%%%%%%%%%%%%%%%%%%%%%%%%%%%%%%%%%%%%%%%%%%%%%%%%%%%%%%%%%%%%%%%%%%

VPMS\,J1342+2840 is an unresolved radio-loud quasar at $z$$\approx$1.3 
with a very peculiar spectrum. Both the optical spectrum and the 
broad-band SED are not consistent with 
obvious reddening by SMC-like dust. Either the reddening curve is 
(a) much steeper between 2000 and 2900\AA\, and (b) flat at longer 
wavelengths, or the absorption comes from unusual BAL features dominated by
wide, overlapping troughs with velocity-dependent 
partial covering of the central source. The dusty scattering 
hypothesis can be tested by 
spectropolarimetry (e.g. Hines et al. \cite{Hin01}).
With its high variability index VPMS\,J1342+2840
is one of the highest priority quasar candidates at the
faint end of the VPMS. Obviously, the VPMS provides a useful tool 
for the detection of unconventional quasars.

\begin{acknowledgements}
We are grateful to the anonymous referee and to Patrick Hall
for helpful comments.
This research is based on observations made with the Italian Telescopio 
Nazionale Galileo (TNG) operated on the island of La Palma by the Centro 
Galileo Galilei of the INAF (Istituto Nazionale di Astrofisica) at the 
Spanish Observatorio del Roque de los Muchachos of the Instituto de 
Astrofisica de Canarias and with the 2.2\,m telescope
of the German-Spanish Astronomical Centre, Calar Alto, Spain.
We acknowledge the staff members of these telescopes for their 
kind assistance.   
HM acknowledges financial support from the European Optical 
Infrared Coordination Network for Astronomy, OPTICON, 
and from the
Deut\-sche For\-schungs\-ge\-mein\-schaft under
grants Me1350/17-1 and Me1350/18-1.
DF received funding by the Cosmo Grid project, funded by the
Irish Higher Education Authority.
MH thanks for grants from the Nordrhein-Westf\"alische Akademie der
Wissenschaften.  
\end{acknowledgements}

%________________________________________________________________

{}

\end{document}